\documentclass[10pt,twocolumn,letterpaper]{article}

\usepackage{cvpr}              

\usepackage{textcomp}
\usepackage{stfloats}
\usepackage{url}
\usepackage{hyperref}

\usepackage{subcaption}
\usepackage{verbatim}
\usepackage{graphicx}
\usepackage{algorithm}
\usepackage{multirow}
\usepackage{booktabs}
\usepackage{color}
\usepackage{textcomp}
\definecolor{cvprblue}{rgb}{0.21,0.49,0.74}

\def\paperID{6284} 
\def\confName{CVPR}
\def\confYear{2026}

\title{PopSim: Social Network Simulation for Social Media Popularity Prediction}

\author{
    Yijun Liu$^{1,2,3}$ \quad Wu Liu$^{4}$\thanks{Corresponding author (liuwu@ustc.edu.cn)} \quad Xiaoyan Gu$^{1,2,3}$ \quad Allen He$^{5}$ \quad Weiping Wang$^{1}$ \quad Yongdong Zhang$^{4}$ \\
    $^{1}$The Institute of Information Engineering, Chinese Academy of Sciences \\
    $^{2}$The School of Cyber Security, University of the Chinese Academy of Sciences \\
    $^{3}$State Key Laboratory of Cyberspace Security Defense \\
    $^{4}$The School of Information Science and Technology, University of Science and Technology of China \\
    $^{5}$JD AI Research \\
}

\begin{document}
\maketitle
\begin{abstract}
Accurately predicting the popularity of user-generated content (UGC) is essential for advancing social media analytics and recommendation systems. 
Existing approaches typically follow an inductive paradigm, where researchers train static models on historical data for popularity prediction.
However, the UGC propagation is inherently a dynamic process, and static modeling based on historical features fails to capture the complex interactions and nonlinear evolution. 
In this paper, we propose \textbf{PopSim}, a \textbf{novel simulation-based paradigm} for social media popularity prediction (SMPP). 
Unlike the inductive paradigm, PopSim leverages the large language models (LLMs)-based multi-agent social network sandbox to simulate UGC propagation dynamics for popularity prediction. 
Specifically, to effectively model the UGC propagation process in the network, we design a social-mean-field-based agent interaction mechanism, which models the dual-channel and bidirectional individual-population interactions, enhancing agents' global perception and decision-making capabilities. 
In addition, we propose a multi-source information aggregation module that transforms heterogeneous social metadata into a uniform formulation for LLMs. Finally, propagation dynamics with multimodal information are fused to provide comprehensive popularity prediction.
Extensive experiments on real-world datasets demonstrate that SimPop consistently outperforms the state-of-the-art methods, reducing prediction error by an average of 8.82\%, offering a new perspective for research on the SMPP task.
\end{abstract}
\section{Introduction}
Social media platforms play an essential role in people’s daily lives. Popularity on social media, defined as the influence of shared online posts, serves as a critical metric for evaluating user engagement and public opinion dynamics. It enables platforms to analyze behavioral trends, forecast the diffusion of information, and detect emerging topics \cite{DBLP:conf/mm/ChenHYCHZZ22,DBLP:conf/aaai/ZhangLCXLZZ24,DBLP:journals/tomccap/XuWLWLZZL25}. Accurate prediction of which posts or topics will become popular is of great significance, as it offers substantial value for online marketing \cite{DBLP:conf/icmcs/GuXTHHZ0G24,DBLP:conf/kdd/LiLMWP15}, recommendation systems \cite{DBLP:conf/cvpr/DongLCPZ24,DBLP:conf/cvpr/LiFYDTS23}, and network traffic management \cite{DBLP:conf/nips/0001YL0020,DBLP:journals/tnsm/DAlconzoDMMC19}.

\begin{figure}[!t]
  \centering
  \includegraphics[width=\linewidth]{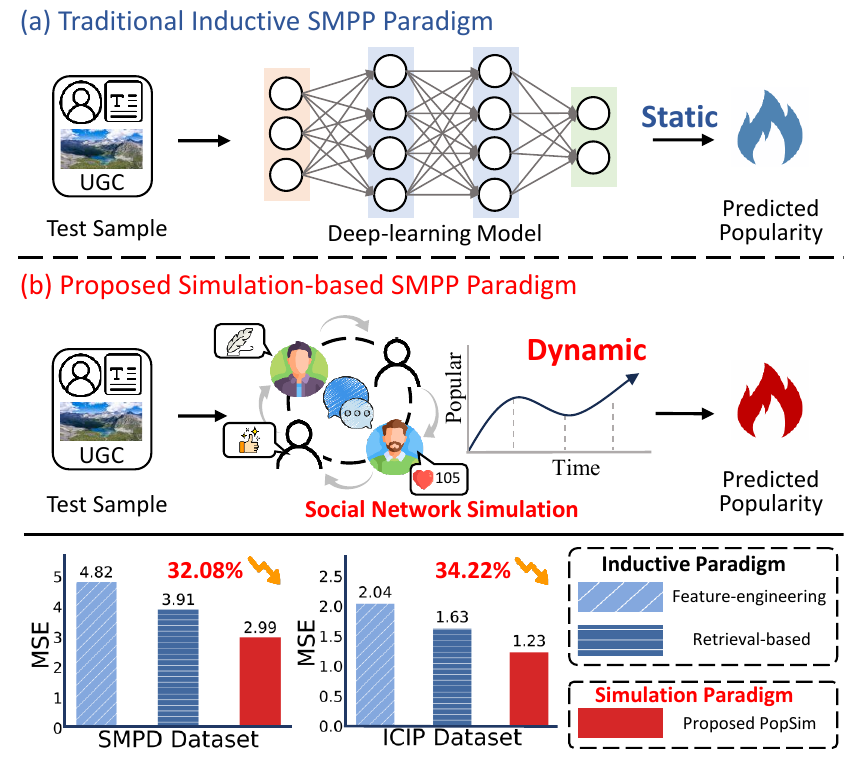}
  \caption{Comparison of different paradigms for the SMPP task. (a) The traditional inductive SMPP paradigm constructs a static popularity prediction model by fitting historical data. (b) The proposed simulation-based paradigm dynamically simulates the UGC propagation for future popularity prediction, which achieves significant improvements.}
  \label{fig:demo}
\end{figure}

Existing Social Media Popularity Prediction (SMPP) research \cite{kang2019catboost,wu2022deeply,xu2020multimodal} mainly follows an inductive paradigm, where researchers learn specific multimodal User-Generated Content (UGC) features from historical data to construct static popularity prediction models, as shown in Fig.~\ref{fig:demo}(a).
Feature-engineering-based methods are the most representative approach among them, these methods extract features from user profiles \cite{DBLP:conf/aaai/WuMCZ16,DBLP:conf/mm/LiHHL18,DBLP:journals/ipm/QianXLLJCL22}, images \cite{DBLP:conf/mir/CappalloMS15,DBLP:conf/www/ZhangWWZ18,DBLP:conf/mm/LaiZZ20}, and time series \cite{DBLP:conf/ijcai/WuCZHLM17,DBLP:conf/mm/WangWCHMZ20,DBLP:conf/mm/TanLLZZ22}, aiming to predict UGC popularity through well-designed functions directly.
The reliance on historical experiences restricts their applicability.
To address this issue, retrieval-based methods have been proposed in \cite{DBLP:conf/mm/JiPRC23,DBLP:conf/sigir/ZhongLZCZ024,DBLP:conf/aaai/XuZ0S25}.
These methods model the relationships among UGC through retrieval-augmented techniques to improve UGC representations. However, determining the propagation trends of UGC remains a major challenge in SMPP.
Typically, highly popular UGC tends to attract widespread user engagement and spreads rapidly, which are crucial factors in determining UGC popularity.
Existing feature-engineering or retrieval-based approaches treat UGC popularity prediction as a static induction based on historical data, while overlooking the dynamic nature and interactive mechanisms of UGC propagation. 

In contrast, the simulation-based paradigm in social science research offers a more dynamic and generalizable framework for social media analytics. 
It facilitates a paradigm shift from inductive interpretation of historical data to deductive exploration of the unknown, enabling a better understanding and prediction of evolving trends in social behavior.
Therefore, as illustrated in Fig.~\ref{fig:demo}(b), we propose to model the UGC propagation dynamics through an LLM-based multi-agent social network sandbox for popularity prediction. The social network simulate how content spreads and resonates across user communities, thereby improving SMPP performance. 
Nevertheless, the simulation still faces several challenges:
(1) given the massive volume of UGC, how to accurately and efficiently simulate the rapidly changing social media propagation process;
(2) how to effectively integrate multi-source information, including images, text, and propagation context, for popularity prediction.

To address these challenges, this paper introduces \textbf{PopSim: a novel simulation-based SMPP paradigm}. 
In contrast to the traditional inductive paradigm, PopSim leverages an LLM-based multi-agent social network sandbox to model UGC propagation dynamics for popularity prediction. 
Rather than relying solely on static UGC features, this paradigm considers the temporal evolution, user interactions, and public opinion trends. 
Specifically, PopSim reformulates the popularity prediction process in a \textit{simulation-and-predict} manner.
In the simulation phase, 
we input UGC into the social network sandbox and model its propagation process to provide rich dynamic UGC propagation features for popularity prediction.
To accurately and efficiently simulate the large-scale social media propagation process, we propose a social-mean-field-based (SMF-based) agent interaction mechanism. 
In this mechanism, we construct a dual-channel social mean field to represent the behaviors and states of all agents in the network. The textual mean field channel takes all agents' behaviors as input to generate the representations of the UGC propagation state. The numerical mean field channel models agents' opinion dynamics on UGC.
Then, each agent interacts with this dual-channel social mean field to perceive the global social network state, thereby enabling more effective decision-making.
After the simulation phase, we can obtain the final social mean field state as the UGC propagation feature for popularity prediction.
In the prediction phase, due to the heterogeneous and discrete nature of UGC text and propagation features, we propose a multi-source information aggregation module.
This module can transform heterogeneous social metadata into a uniform formulation, facilitating a more nuanced understanding by LLMs.
Finally, we train a prediction model to analyze multimodal UGC content and propagation dynamics for popularity prediction.

To evaluate the effectiveness of PopSim, we conduct extensive experiments on two large-scale real-world datasets. Experimental results demonstrate that PopSim outperforms existing state-of-the-art models across all metrics, reducing prediction error by an average of 8.82\%. Further ablation studies validate the effectiveness and robustness of PopSim.

In summary, our main contributions are as follows:
\begin{itemize}
    \item We introduce PopSim, a novel simulation-based SMPP paradigm. Unlike the traditional inductive paradigm, PopSim leverages LLM-based agents to construct a social network sandbox for modeling UGC propagation dynamics, enabling more accurate popularity prediction. 
    \item We propose an SMF-based agent interaction mechanism, which models the dual-channel and bidirectional interactions between individuals and population, thereby facilitating more effective UGC propagation simulation.
    \item Extensive experiments on two SMPP datasets demonstrate the effectiveness of PopSim, with an average reduction of 8.82\% in prediction loss compared to the state-of-the-art models.
\end{itemize}

\begin{figure*}[!t]
  \centering
  \includegraphics[width=\linewidth]{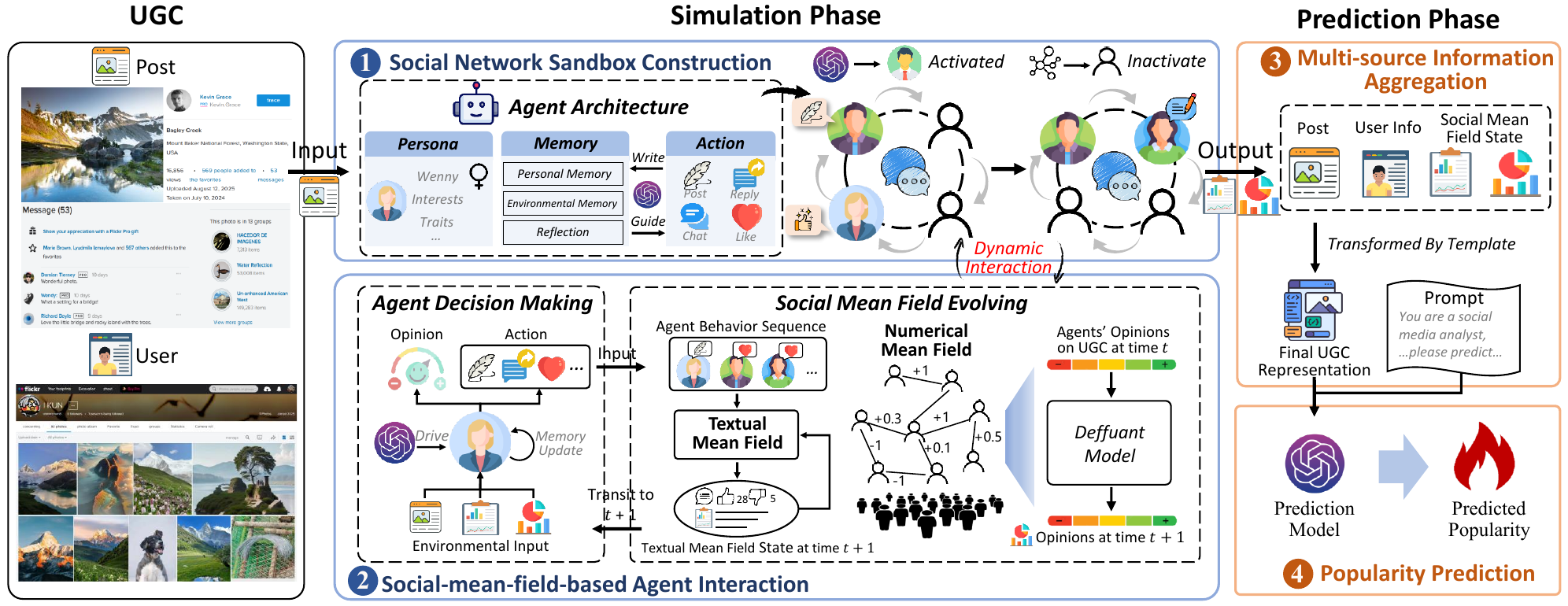}
  \caption{The framework of the PopSim. PopSim reformulates SMPP in a simulation-and-predict manner. In the simulation phase: (1) we construct a social network sandbox with LLM-based multi-agents to simulate social media users; (2) the agents interact through the proposed social mean field mechanism to model the UGC propagation process. The resulting social mean field state is used as UGC propagation features for popularity prediction. In the prediction phase: (3) multi-source information aggregation is developed to transform heterogeneous social media metadata and UGC propagation features into semantically unified formulations; (4) the prediction model then analyzes the above information for popularity prediction.
  }
  \label{fig:overview}
\end{figure*}

\section{Related Work}
\subsection{Social Media Popularity Prediction}
Predicting the popularity of UGC on social media is a key challenge for various social and recommendation applications \cite{DBLP:journals/tomccap/XuWLWLZZL25,DBLP:conf/cvpr/McKeeSSR23,DBLP:journals/pami/DongSLG25}. 
Existing SMPP methods \cite{DBLP:conf/aaai/WuMCZ16,DBLP:conf/mm/LiHHL18,DBLP:conf/mm/LaiZZ20} primarily follow an inductive paradigm, where researchers learn specific multimodal UGC features from historical data to construct static popularity prediction models.
Feature-engineering-based methods are the most representative approach among them, these methods extract features from user profiles \cite{DBLP:conf/aaai/WuMCZ16,DBLP:conf/mm/LiHHL18,DBLP:journals/ipm/QianXLLJCL22}, images \cite{DBLP:conf/mir/CappalloMS15,DBLP:conf/www/ZhangWWZ18,DBLP:conf/mm/LaiZZ20}, and time series \cite{DBLP:conf/ijcai/WuCZHLM17,DBLP:conf/mm/WangWCHMZ20,DBLP:conf/mm/TanLLZZ22}, aiming to predict UGC popularity through well-designed functions directly.
The key limitation of these methods lies in their reliance on historical experiences to learn static UGC representations, which restricts their applicability.
To overcome it, retrieval-based methods \cite{feng2019hypergraph,DBLP:conf/kdd/JiLLSL0023,DBLP:conf/mm/JiPRC23,DBLP:conf/sigir/ZhongLZCZ024,DBLP:conf/aaai/XuZ0S25} model UGC relationships through retrieval-augmented techniques and integrate multi-source features. 
For instance, HGNN \cite{feng2019hypergraph} uses hypergraph structures and graph convolutions to model higher-order UGC correlations. 
RAGTrans \cite{DBLP:conf/kdd/ChengZXTZ024} enhances UGC representations through neighborhood knowledge aggregation in a knowledge-enhanced hypergraph. 
SKAPP \cite{DBLP:conf/aaai/XuZ0S25} introduces a meta-retrieval strategy that incorporates multimodal content semantics to retrieve relevant UGC knowledge, enhancing UGC representations and achieving state-of-the-art performance.
However, the essence of these methods lies in static induction of historical data, which fails to capture the dynamic interactions and evolution of UGC propagation, thus limiting their performance.

\subsection{LLM-agent-based Social Network Simulation}
With the significant development of LLMs, LLM-agent-based social networks simulation has recently attracted widespread attention \cite{DBLP:conf/acl/MouWH24,DBLP:journals/corr/abs-2504-10157,DBLP:conf/cvpr/LiX0CJN25}. 
These methods leverage LLM-based agents to simulate the dynamic interactions of diverse individuals in social networks, offering valuable insights for predicting and analyzing potential real-world events
\cite{DBLP:conf/uist/ParkOCMLB23,DBLP:journals/corr/abs-2412-09237,DBLP:conf/ijcai/LiuCZG0024}.
For example, GenerativeAgent \cite{DBLP:conf/uist/ParkOCMLB23} introduces an AI town constructed by dozens of LLM-based agents and studies the various spontaneous behaviors of these agents in a sandbox environment. 
Hisim \cite{DBLP:conf/acl/MouWH24} uses an LLM-based agent system to study the evolution of events in social networks.
Furthermore, as small-scale simulations fail to capture complex emergent social behaviors, some works \cite{DBLP:journals/corr/abs-2412-09237,DBLP:journals/corr/abs-2504-10157,liu2025rumorsphere,DBLP:journals/corr/abs-2410-04360} propose to scale up the simulation. 
For instance, LMAgent \cite{DBLP:journals/corr/abs-2412-09237} designs a multi-user simulation system with tens of thousands of users in e-commerce scenarios, providing an effective method for studying large-scale consumer behavior. 
SocioVerse \cite{DBLP:journals/corr/abs-2504-10157} utilizes four powerful alignment components and a ten-million-level user pool to achieve precise large-scale simulation on social, political, and economic scenarios.
These applications highlight the effectiveness of LLM-based multi-agent systems in advancing social media analytics and social science research.

In this work, we innovatively combine LLM-agent-based social network simulation with the SMPP task and propose a social-mean-field-based agent interaction mechanism to model the temporal evolution, user interactions, and public opinion trends of UGC propagation, providing a novel paradigm for SMPP.

\section{Methodology}
\subsection{Overview}
As shown in Fig.~\ref{fig:overview}, the proposed PopSim reformulates the SMPP in a \textit{simulation-and-predict} manner. 
In the simulation phase, as shown in Fig.~\ref{fig:overview} (1), we construct a social network sandbox to model the UGC propagation process. It includes up to 1,000 LLM-based agents, each with independent profile, memory, and action modules, activated adaptively based on social participation inequality rules \cite{socialparti} to simulate social media user behavior. 
To efficiently model the large-scale and dynamic UGC propagation process, we propose a social-mean-field-based agent interaction mechanism, as shown in Fig.~\ref{fig:overview} (2). 
This mechanism utilizes mean field theory \cite{lasry2007mean} to maintain both textual and numerical social mean fields to encode population-centric, evolving social network state representations. 
The textual mean field channel takes all agents' behaviors as input, using LLMs to generate the representations of the UGC propagation state. The numerical mean field channel, based on the Deffuant model \cite{DBLP:journals/advcs/DeffuantNAW00}, models agents' opinion dynamics on UGC.
Then, agents make decisions and interact based on the social mean field state, and their behaviors, in turn, influence the update of the social mean field.
We use the social mean field state as propagation features for subsequent popularity prediction.
In the prediction phase,  as shown in Fig.~\ref{fig:overview} (3) and (4), we trained a prediction model to comprehensively analyze multimodal UGC content and propagation features for popularity prediction.
Due to the heterogeneous and discrete nature of UGC text and propagation features, we employ a multi-source aggregation module to transform social metadata and UGC propagation features into a semantically rich formulation for unified feature processing.

\subsection{Social Network Sandbox Construction}
The propagation of UGC and its potential popularity are influenced by various factors such as user interests and social trends. 
Traditional signal-based models overlook the complexity of the real world, thereby failing to comprehensively model UGC propagation.
Therefore, in the simulation phase, we propose constructing an LLM-based multi-agent social network sandbox, leveraging the powerful understanding and reasoning capabilities of LLMs to model the dynamic UGC propagation for predicting its future popularity. The sandbox construction involves the design of the agent architecture and the sandbox environment.

\subsubsection{Agent Architecture}
To model social media user behavior at a fine-grained level, we design an agent architecture based on LLM. 
The core features of this architecture are inherited from RumorSphere \cite{liu2025rumorsphere}, and it consists of the persona, memory, and action modules. 
The persona module includes demographic attributes such as name, gender, occupation, interests, and personality traits, which are closely related to the agent’s potential position in social interactions \cite{DBLP:conf/hicss/BrunkerWMM20}. 
The memory module is used to record detailed information about the UGC received by the agent (e.g., post content, comment content, or messages from friends) as well as its interactions with the environment or other agents, guiding the agent’s subsequent behavior. 
The action module supports five common behaviors in the Twitter environment: (1) \textit{post}: sharing original content with friends; (2) \textit{retweet}: forwarding existing tweets; (3) \textit{reply}: responding to tweets; (4) \textit{like}: endorsing tweets; (5) \textit{do nothing}: remaining silent. 
Through these behaviors, the agent continuously interacts, providing a foundation for simulating the evolution of UGC propagation in social networks. 
Furthermore, based on the social participation inequality rules \cite{socialparti}, we design an agent adaptive activation module that uses the similarity between agent profiles and UGC to dynamically calculate the activation probability of agents, reducing computational complexity while making the simulation more aligned with real-world information propagation. 
The details of agent architecture and initialization are presented in Appendix A.

\subsubsection{Sandbox Environment}
In this sandbox environment, agents adaptively activate and take turns acting, interacting based on a social mean-field mechanism (described in Section~\ref{sec:smf}), and freely engaging in social interactions. UGC is introduced into this sandbox, and the agents' social behaviors facilitate opinion exchange and the propagation of UGC. 
We use the UGC propagation state representation for subsequent popularity prediction.
More simulation details are presented in Appendix B.

\subsection{Social-mean-field-based Agent Interaction}
\label{sec:smf}
UGC propagation is driven by user interactions on social media.
As the population size and time horizon grow, modeling all agents' perceptions and interactions becomes challenging. 
To address this, we introduce a social-mean-field-based agent interaction mechanism.
Based on mean field theory \cite{lasry2007mean}, we construct textual and numerical mean fields to model population-centric, evolving social network state representations. 
Agents make decisions and interact based on the social mean field state, and their behaviors, in turn, influence the update of the social mean field.
Through these dual-channel and bidirectional interactions, we can effectively model the UGC propagation process.
The agent interaction focuses primarily on two aspects: social mean field evolving and agent decision-making.

\subsubsection{Social Mean Field Evolving}
The social mean field $\mathcal{M}$ consists of a textual mean field channel $\mathcal{M}_{text}$ and a numerical mean field channel $\mathcal{M}_{num}$.
The textual mean field takes all agents' behaviors as input, using LLMs to generate textual representations of the UGC propagation state, capturing emergent social behaviors and providing strong flexibility and adaptability. 
The numerical mean field, based on the Deffuant model \cite{DBLP:journals/advcs/DeffuantNAW00}, models agents' opinion dynamics on UGC, offering a more accurate depiction of agent interactions. 
The numerical mean-field state represents the average opinion score of all agents in the social network towards UGC, providing key information for popularity prediction.

Formally, consider that at time $t$, each agent's behavior is denoted as $\{a^{t}_1, a^{t}_2, ...\}$. 
We utilize LLM to summarize the UGC propagation state at time $t$, i.e., the textual mean field state $\mathcal{M}_{text}^{t}$:
\begin{equation}
M_{\text{text}}^{t} = f_{sum}(\{T_{a^{t}_1},T_{a^{t}_2},...\},M_{\text{text}}^{t-1},\mathcal{P}_{text}),
\end{equation}
where $f_{sum}$ is the prompt function to guide the LLM in summarizing the UGC propagation state, and $\mathcal{P}_{text}$ is the prompt words. $T_{a^{t}_i}$ denotes the textualized agent behavior descriptions obtained through a fixed template transformation. The details of the transformation template and prompts are presented in Appendix C.

For the numerical mean field, we use the Deffuant model to simulate the opinion dynamics of agents on UGC within the social network. This model captures the mutual influence process between agents. 
In each interaction round, agents update their opinions on UGC (ranging from 0 to 10, where 10 indicates high interest and 0 indicates no interest) according to the following equation:
\begin{equation}
    \Delta o_i^t= o_i^{t+1}-o_i^{t}= f_{update}(o_i^{t},\mathcal{J}_{i,t}),
\end{equation}
\begin{equation}
f_{update}(o_i^{t},\mathcal{J}_{i,t}) =\frac{1}{|\mathcal{J}_{i,t}|}\sum_{o_j^t\in \mathcal{J}_{i,t}} \alpha_j \cdot \left(o_j^t-o_i^t\right),
\end{equation}
where $f_{update}$ is the opinion update function, $o_i^t$ is the opinion state of agent $\textbf{a}_i$ on UGC at time $t$, $\Delta o_i^t$ is the opinion change of $\textbf{a}_i$ from
time step $t$ to $t + 1$, $\alpha_j$ is the influence weight of agent $\textbf{a}_j$.
$\mathcal{J}_{i,t}$ is the set of agents that will exert influence on $\textbf{a}_i$:
\begin{align}
    \mathcal{J}_{i,t} = \{o_j^t \mid j  \in \mathcal{N}(i) , j \neq i, \, |o_j^t - o_i^{t}| < \varepsilon\},
\end{align}
where $\mathcal{N}(i)$ is the set of neighboring agents of agent $\textbf{a}_i$, $\varepsilon$ is a threshold that limits the acceptable difference between $o_j^t$ and $o_i^{t}$ for influence.
This implies that agent $j$ exerts an assimilative effect on agent $i$ only when their opinions on UGC are sufficiently close.

Finally, we average the opinion of all agents to obtain the numerical state of the entire social network towards UGC:
\begin{equation}
M_{\text{num}}^{t} = \frac{1}{N}\sum_{i=1}^{N}o_i^t,
\end{equation}
$M_{\text{num}}$ records the evolution of the agents' opinion scores towards UGC at each moment during its propagation, providing crucial insights for popularity prediction.

\subsubsection{Agent Decision Making}
The social mean-field state is determined by the behavior of all agents and, in turn, serves as environmental input that influences the agents' decisions.
Formally, the action $a_i^{t+1}$ taken by agent $\textbf{a}_i$ at time $t+1$ is determined jointly by its current opinion state $o^t_i$ on UGC, the memory $\mathbb{M}^t_i$, the UGC $c_u$, and the social mean field state $\mathcal{M}^t$:
\begin{equation}
    a^{t+1}_i, o_i^{t+1} = f_{decision}(c_u, o^t_i,\mathbb{M}^t_i, \mathcal{M}^t,\mathcal{P}_{action}),
\end{equation}
where $f_{decision}$ is the prompt function that uses LLM to infer the agent's next behavior and estimate the opinion score, $\mathcal{P}_{action}$ represents the prompt words, and $\mathbb{M}^t_i=\{m^1_i,m^2_i,...\}$, denotes the memory entries, where $m^j_i$ records the agent's historical behavior.
Full examples can be found in Appendix C.

Through continuous interactions, we can model the UGC propagation process in the social network sandbox. 
The textual mean field represents the textual UGC propagation state, while the numerical mean field records the evolution of agent opinion scores towards the UGC at each moment. 
They serve as UGC propagation features, providing crucial insights for subsequent popularity prediction.

\subsection{Multi-source Information Aggregation}
The social media metadata and the UGC propagation features from the simulation phase are typically heterogeneous and stored in structured formats, making them unsuitable for direct use in popularity prediction.
Therefore, in the prediction phase, to facilitate unified feature processing and enhance the representation, we transform these multi-source data into semantically enriched UGC text. 
Specifically, we use specific templates to expand the field names in the metadata and merge them with their corresponding field values, transforming all text information into coherent, semantically rich text $r_u$. 
This process provides enhanced context and clarity for subsequent LLM-based popularity predictions. The templates for aggregating multi-source information are presented in Appendix D.

\subsection{Popularity Prediction}
Finally, we trained a prediction model to comprehensively analyze multimodal UGC content and propagation features for popularity prediction:
\begin{equation}
    \hat{y} = \pi_p(r_u, I,\mathcal{P}_{prediction}),
\end{equation}
where $\pi_p$ denotes the multimodal LLM-based prediction model, $r_u$ refers to the semantically enriched UGC text, $I$ denotes the image in UGC, and $\mathcal{P}_{prediction}$ is the prompt that guides the prediction model in predicting popularity based on these inputs. 
The training objective is to minimize the difference between the model's predicted popularity scores $\hat{y}_i$ and the ground truth $y_i$, achieved through the cross-entropy loss function:
\begin{equation}
    \mathcal{L}_{\mathrm{CE}}=-\frac{1}{N_{train}} \sum_{i=1}^{N_{train}}\left[y_i \log \hat{y}_i+\left(1-y_i\right) \log \left(1-\hat{y}_i\right)\right],
\end{equation}
where $N_{train}$ is the number of training samples.

\begin{table*}[!t]
\centering
\caption{Results of social media popularity prediction performance. Best (bold) and \underline{second best} numbers are highlighted in each column.
All experiments were conducted on the dataset with temporal splits, as random splits may pose a risk of label leakage.
}
\label{tab:results}
\begin{tabular}{ccccccccc}
\toprule
                                  &                                        &                                 & \multicolumn{3}{c}{\textbf{SMPD}}                                           & \multicolumn{3}{c}{\textbf{ICIP}}                                           \\ \cline{4-9} 
\multirow{-2}{*}{\textbf{Method}} & \multirow{-2}{*}{\textbf{Publication}} & \multirow{-2}{*}{\textbf{Type}} & MAE                     & MSE                     & SRC                     & MAE                     & MSE                     & SRC                     \\
\midrule
SVR \cite{DBLP:conf/www/KhoslaSH14}                               & WWW 2014                               &                                 & 2.0208                  & 6.2996                  & 0.2163                  &  0.8941                 & 1.9009                  & 0.5241                  \\
UHAN \cite{DBLP:conf/www/ZhangWWZ18}                               & WWW 2018                               &                                 & 1.4833                  &  3.8471                  & 0.5541                  &  1.2824                 & 2.7492                  &  0.3981                  \\
HyFea \cite{DBLP:conf/mm/LaiZZ20}                             & MM 2020                                &                                 &  1.7080                 & 4.7429                   & 0.4677                  & 1.0181                  & 1.9013                  & 0.4497                  \\
BLIP \cite{DBLP:conf/icml/0001LXH22}                              & ICML 2022                              &                                 &  1.6340                 & 4.3884                   & 0.5269                  & 0.9961                  & 2.0646                  & 0.3603                  \\
JAB \cite{weissburg2022judging}                              & AAAI 2022                                &                                 &  1.9359                 &  6.1882                  & 0.2353                  & 0.9289                  & 1.8606                  & 0.3057                  \\
MFTM \cite{DBLP:conf/mm/HsuLHT23}                              & MM 2023                                &                                 & 1.5481                  & 4.0222                  & 0.5849                  & 0.9772                  & 1.8970                  & 0.4156                  \\
HMMVED \cite{DBLP:journals/tmm/XieZC23}                           & TMM 2023                               & \multirow{-9}{*}{\begin{tabular}[c]{@{}c@{}}Feature-\\ engineering\end{tabular}}       & 1.3636                  & 3.7154                  & 0.6352                  & 0.9497                  & 1.8556                  & 0.4524                  \\
\midrule
HGNN \cite{feng2019hypergraph}                              & AAAI 2019                                &                                 &  1.6061                  & 5.1770                  &  0.4371                  & 0.9093                  & 1.6711                  & 0.4423                  \\
NIPA \cite{DBLP:conf/mm/JiPRC23}                              & MM 2023                                &                                 & 1.6532                  & 4.2538                  & 0.4086                  & 0.9980                  & 1.9999                  & 0.3989                  \\
NMRA \cite{DBLP:conf/sigir/ZhongLZCZ024}                              & SIGIR 2024                             &                                 &  1.3730                 & 3.5119                   & 0.6423                  & 0.8684                  & 1.7600                  & 0.5439                  \\
RAGTrans \cite{DBLP:conf/kdd/ChengZXTZ024}                         & KDD 2024                               &                                 & 1.3396                  & \underline{3.2763}                  & 0.5859                  &  \underline{0.7149}                  & \underline{1.2351}                  & \underline{0.5914}                  \\
SKAPP  \cite{DBLP:conf/aaai/XuZ0S25}                            & AAAI 2025                              & \multirow{-5}{*}{Retrieval}     & \underline{1.3365}
                  & 3.3573
                  & \underline{0.6828}
                  & 0.7367                 & 1.4662                  & 0.5765                  \\
\midrule
Llama                              & -                                      & LLM                      & 1.5926              & 3.9749
                  & 0.6365
                  & 0.8722                  & 1.8263                  & 0.5387                  \\
\midrule
Ours                              & -                                      & Simulation                      & \textbf{1.2989}              & \textbf{2.9872}
                  & \textbf{0.7026}
                  & \textbf{0.7008}                  & \textbf{1.2257}                  & \textbf{0.6192}                  \\
\bottomrule
\end{tabular}
\end{table*}

\begin{table}[!t]
\centering
\caption{Dataset statistics.}
\label{tab:data}
\resizebox{\linewidth}{!}{
\begin{tabular}{c|cccc}
\toprule
\textbf{Dataset} & \textbf{\# UGCs} & \textbf{\# Users} &  \textbf{\#Tags} & \textbf{avg. Text Length} \\
\midrule
\textbf{ICIP} \cite{DBLP:conf/iciap/OrtisFB19}    & 20,337           & 17,302                                & 18,844          & 27.68                     \\
\textbf{SMPD} \cite{wu2019smp}   & 305,613          & 38,312                              & 250,000            & 91.75               \\
\bottomrule
\end{tabular}}
\end{table}

\section{Experiments}
Our experiments span two real-world social media UGC datasets, including comparative experiments (Section~\ref{sec:main}), ablation studies (Section~\ref{sec:abl}), agent behavior analysis (Section~\ref{sec:aba}), simulation scale and rounds analysis (Sections~\ref{sec:ias}--\ref{sec:iir}), efficiency analysis (Sections~\ref{sec:ea}), and a case study (Section~\ref{sec:cs}). More additional experiments and details are provided in Appendix E.

\subsection{Experimental Settings}
\textbf{Datasets:} We evaluate the proposed model using two widely used real-world social media datasets: ICIP \cite{DBLP:conf/iciap/OrtisFB19} and SMPD \cite{wu2019smp}. Both datasets contain posts associated with various types of information, including images, categories, tags, user profiles, geolocation, and other attributes. 
The datasets are split chronologically into training, validation, and test sets in an 8:1:1 ratio. Table~\ref{tab:data} presents the basic statistics of the datasets.

\noindent
\textbf{Metrics:} Following existing works \cite{wu2019smp,DBLP:conf/aaai/XuZ0S25,zhang2024contrastive}, we use mean squared error (MSE), mean absolute error (MAE), and Spearman’s rank correlation (SRC) as standard evaluation metrics. 
The details of these metrics are presented in Appendix E.1.
\noindent
\textbf{Baselines:} We compare PopSim with the following baselines. They can be roughly divided into feature-engineering approaches: SVR \cite{DBLP:conf/www/KhoslaSH14}, UHAN \cite{DBLP:conf/www/ZhangWWZ18}, HyFea \cite{DBLP:conf/mm/LaiZZ20}, BLIP \cite{DBLP:conf/icml/0001LXH22}, JAB\cite{weissburg2022judging}, MFTM \cite{DBLP:conf/mm/HsuLHT23}, HMMVED \cite{DBLP:journals/tmm/XieZC23}; and retrieval-based approaches: HGNN \cite{feng2019hypergraph}, NIPA \cite{DBLP:conf/mm/JiPRC23}, NMRA \cite{DBLP:conf/sigir/ZhongLZCZ024},  RAGTrans \cite{DBLP:conf/kdd/ChengZXTZ024}, SKAPP \cite{DBLP:conf/aaai/XuZ0S25}.
We also use the Llama model (Llama-3.2-11B-Vision) \cite{touvron2023llama} to directly predict popularity as an LLM-based baseline for comparison.

\noindent
\textbf{Implementation:}
In our experiment, we use Llama3.1-8B as the agent's engine, and Llama-3.2-11B-Vision \cite{DBLP:journals/corr/abs-2407-21783} as the prediction model. The random seed is set to 1 for reproducibility, with the temperature set to 1.0 to ensure output diversity. 
In the numerical mean field, the weights of activated and inactive agents are set to 0.8 and 0.2, respectively, with an interaction threshold $\varepsilon$ of 6.

\subsection{Main Results}
\label{sec:main}
Table~\ref{tab:results} presents the performance of PopSim compared to baseline models, with the following observations:
(1) PopSim consistently outperforms all competing baselines across all evaluation metrics on both the SMPD and ICIP datasets. Notably, on the SMPD dataset, it achieves substantial improvements of 2.81\% in MAE, 8.82\% in MSE, and 2.90\% in SRC, compared to state-of-the-art models. These results underscore its effectiveness as a novel simulation-based paradigm for SMPP tasks.
(2) Through social network simulation, PopSim can effectively model UGC propagation trends, providing a more comprehensive and dynamic UGC representation. Compared to the optimal feature-engineering-based and retrieval-based methods, PopSim achieves an average SRC improvement of 10.61\% and 2.90\%, respectively.
(3) PopSim can effectively integrate multi-source heterogeneous information, comprehensively analyzing multimodal UGC content and propagation context. Compared to directly predicting popularity using the LLM, it improves MAE, MSE, and SRC by 18.44\%, 24.85\%, and 10.38\%, respectively.
These findings highlight the effectiveness of the proposed model, presenting a novel paradigm for SMPP and other social media analysis tasks.

\subsection{Ablation Study}
\label{sec:abl}
To investigate the contributions of PopSim’s key modules, we conduct ablation studies on the SMPD dataset. 
The results are shown in Table~\ref{tab:abl}.

\textbf{Effect of UGC propagation simulation.}
To verify the effectiveness of propagation simulation for SMPP, we implement a standard simulation method (Sim), where agents interact with all their neighbors to capture propagation dynamics for popularity prediction. As shown in Table~\ref{tab:abl}, ``Sim'' achieves improvements of 5.48\% in SRC and 7.73\% in MSE, demonstrating that modeling UGC propagation dynamics can effectively improve SMPP performance.

\textbf{Effect of SMF interaction.}
We validated the effectiveness of the SMF-based agent interaction by comparing it with the ``Sim''. As shown in Table~\ref{tab:abl}, ``SMF-T'' and ``SMF-N'' improve SRC by 2.20\% and 2.98\%, respectively, compared to ``Sim''.
This demonstrates that by interacting with the dual-channel social mean field, agents can better perceive environment information and make more effective decisions, thereby facilitating more effective propagation simulation and enhancing SMPP performance.

\textbf{Effect of multi-source information aggregation.}
We investigate the impact of multi-source information aggregation, using the direct concatenation of UGC metadata as a baseline. As shown in Table~\ref{tab:abl}, ``MIA'' improved SRC by 1.01\% and MSE by 3.46\%. This demonstrates that it provides LLMs with semantically richer text, enhancing their understanding and reasoning capabilities, and further improving SMPP performance.

\begin{table}[!t]
\centering
\caption{Ablation study of PopSim on the SMPD dataset. ``Sim'' indicates standard propagation simulation, ``SMF-T'' and ``SMF-N''  indicate textual/numerical social-mean-field-based agent interaction, and ``MIA'' indicates multi-source information aggregation.}
\label{tab:abl}
\resizebox{\linewidth}{!}{
\begin{tabular}{cccccccc}
\toprule
\textbf{Sim} & \textbf{SMF-T} & \textbf{SMF-N} & \textbf{MIA} & \textbf{MAE}    & \textbf{MSE}    & \textbf{SRC}    & \textbf{Improv}  \\
\midrule
-            & -              & -              & -            & 1.5926          & 3.9749          & 0.6365          & -                \\
\midrule
\checkmark            & -              & -              & -            & 1.4274          & 3.6677          & 0.6714          & 5.48\%           \\
\checkmark            & \checkmark              & -              & -            & 1.3361          & 3.2640          & 0.6862          & 7.81\%           \\
\checkmark            & -              & \checkmark              & -            & 1.3182          & 3.1513          & 0.6914          & 8.67\%           \\
\checkmark            & \checkmark              & \checkmark              & -            & 1.3007          & 3.0942          & 0.6956          & 9.76\%           \\
\checkmark            & \checkmark              & \checkmark              & \checkmark            & \textbf{1.2989} & 
\textbf{2.9872} & \textbf{0.7026} & \textbf{10.38\%} \\
\bottomrule
\end{tabular}}
\end{table}

\begin{figure}[!t]
  \centering
    \subfloat[Agent opinion scores.]{\includegraphics[width=0.48\linewidth]{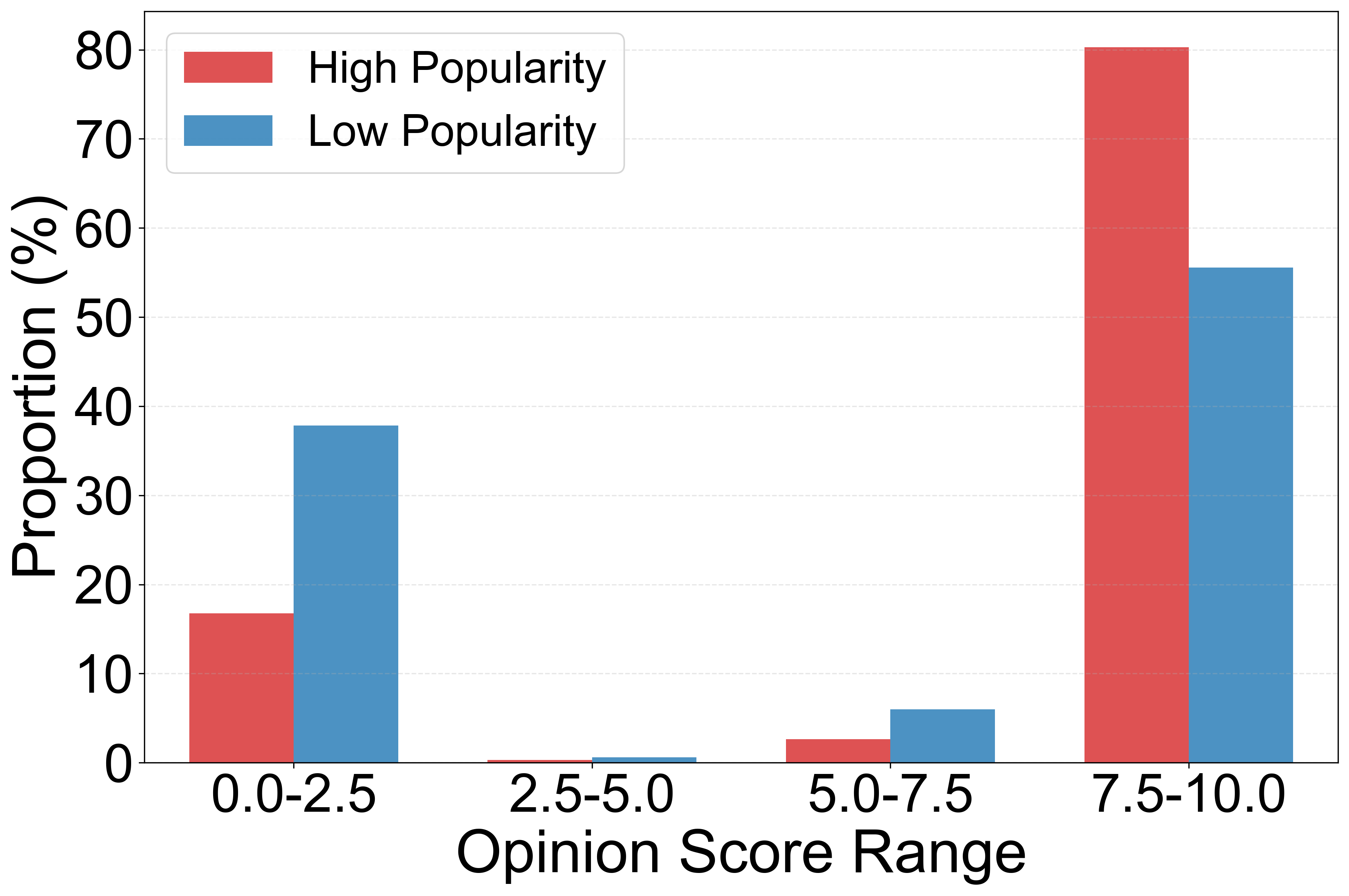}}
    \subfloat[Agent behaviors.]{\includegraphics[width=0.48\linewidth]{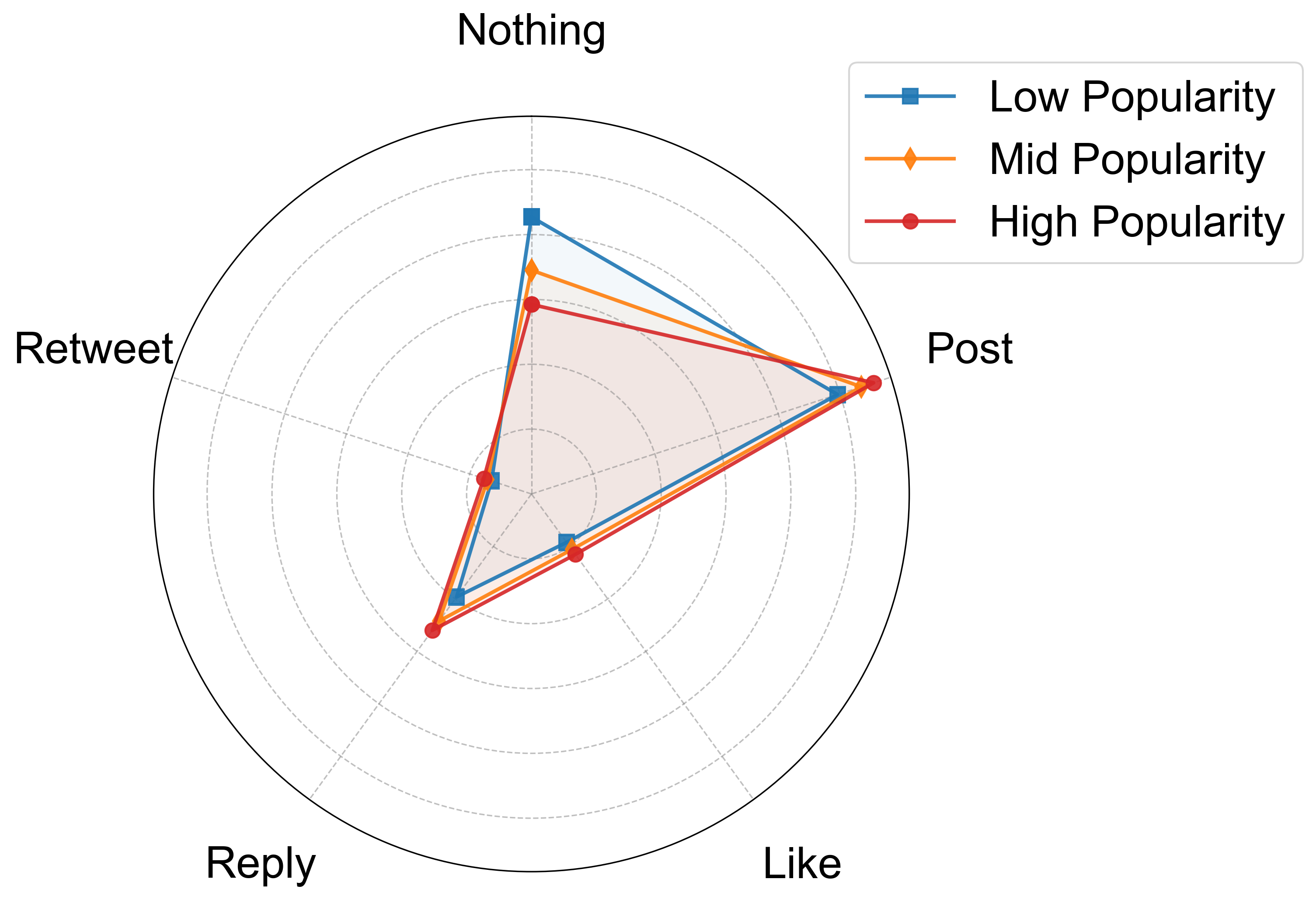}}
  \caption{The distribution of agent behavior and opinion scores under posts with different popularity levels.}
  \label{fig:beha}
\end{figure}

\begin{figure}[!t]
  \centering
    \subfloat[Impact of Agent Scale.]{\includegraphics[width=0.48\linewidth]{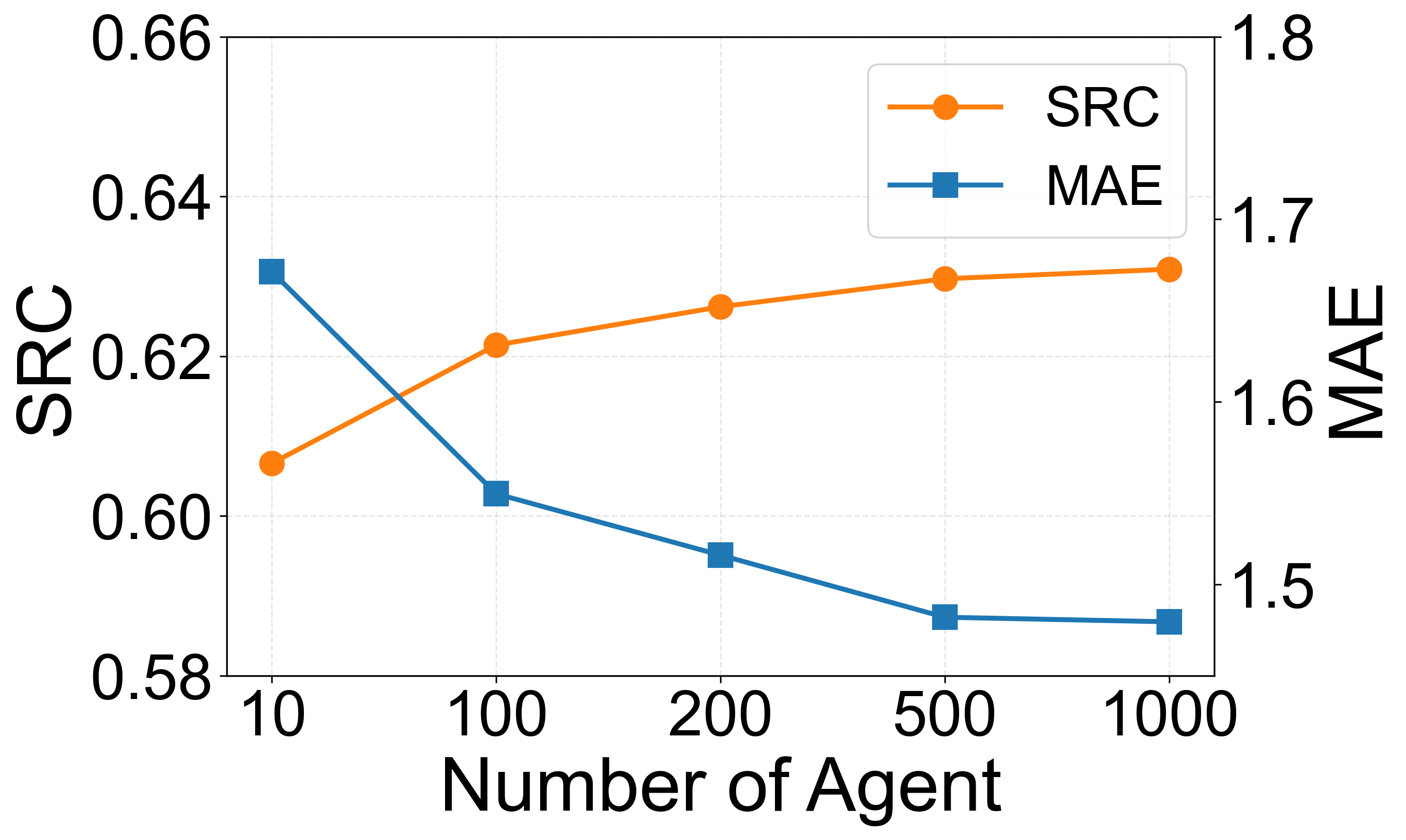}}
    \subfloat[Impact of Interaction Rounds.]{\includegraphics[width=0.48\linewidth]{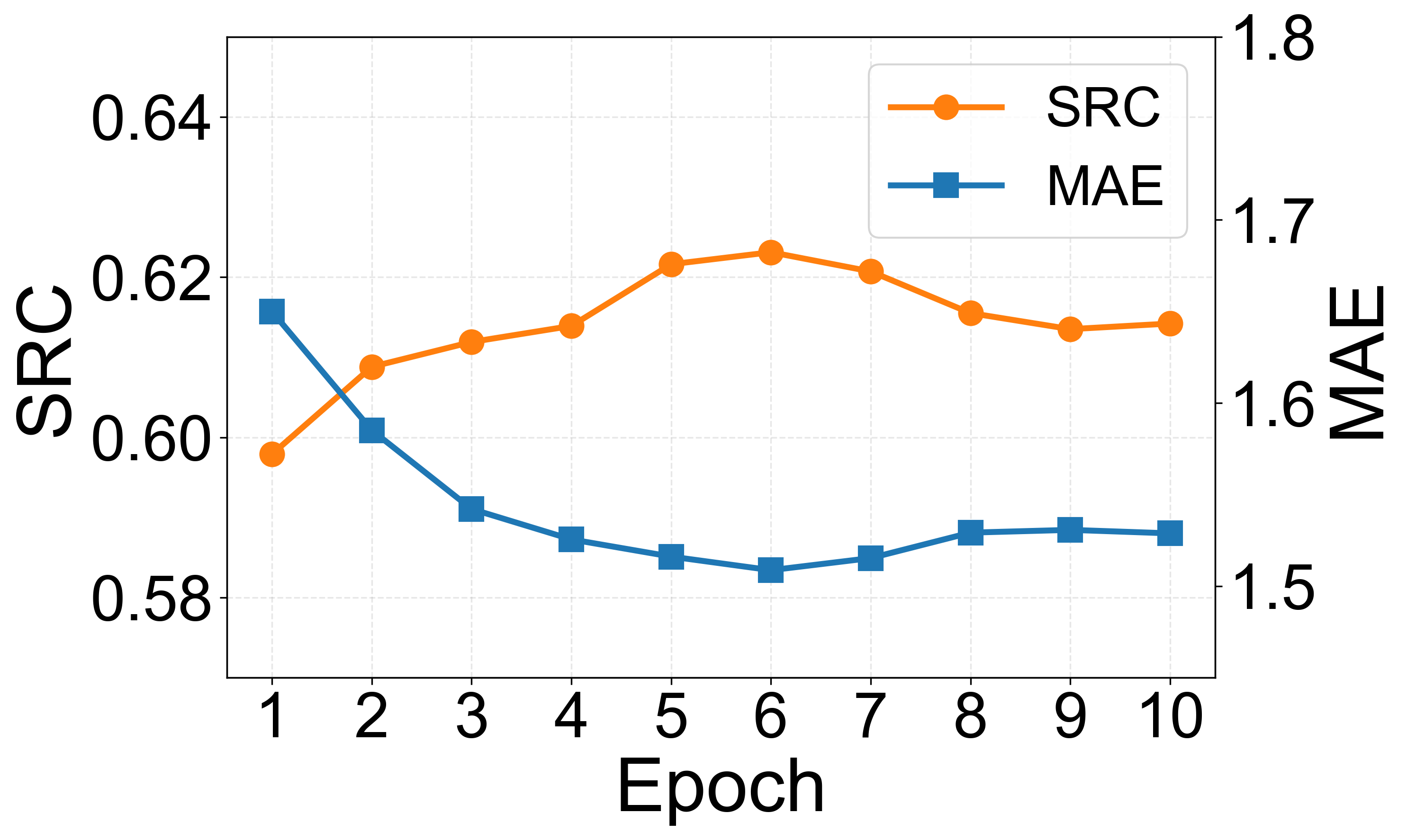}}
  \caption{Model performance under different agent scales and interaction rounds.}
  \label{fig:impact}
\end{figure}

\begin{figure*}[!t]
  \centering
  \includegraphics[width=\linewidth]{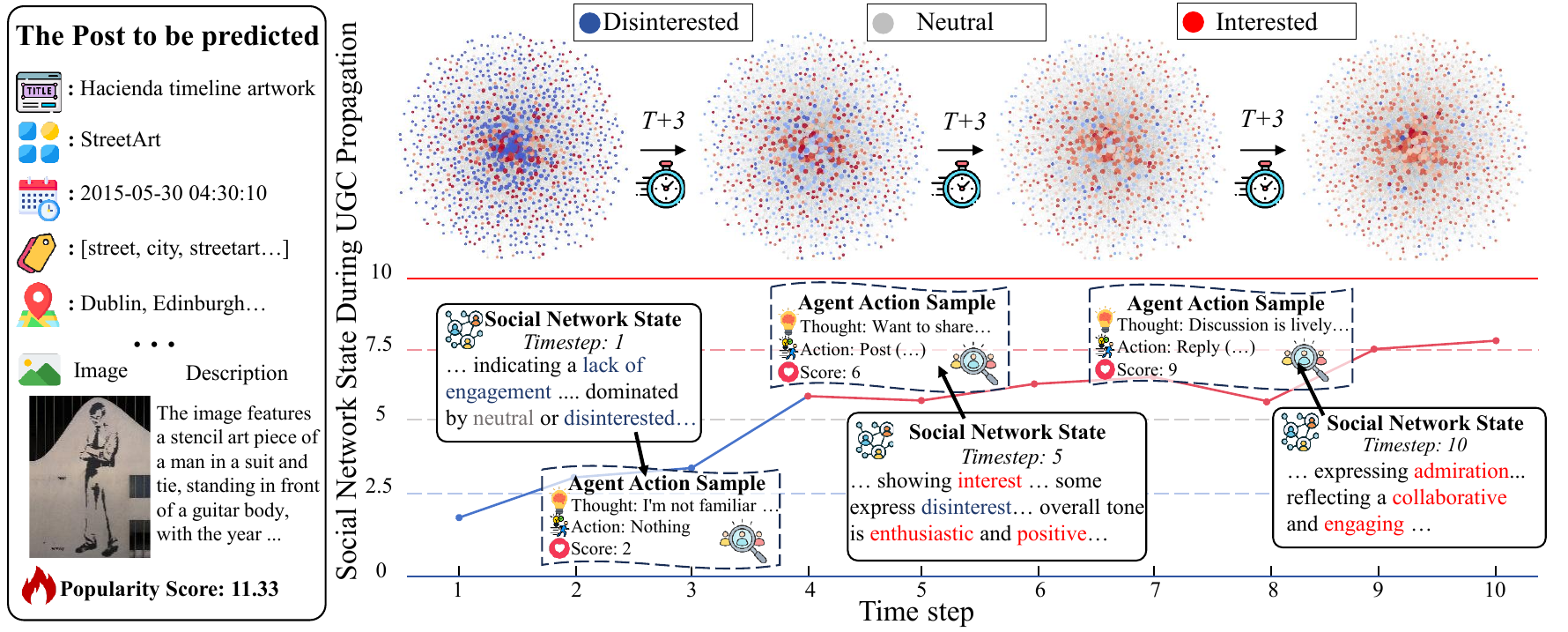}
  \caption{A case study of Popsim. The left shows the post content to be predicted, the upper-right visualizes the opinion scores of all agents in the social network, and the lower-right displays the social mean field state. The line graph represents the dynamics of the numerical mean field, the solid box indicates the textual mean field state, and the dashed box shows typical agent behavior at the current time step.}
  \label{fig:case}
\end{figure*}

\subsection{Agent Behaviors Analysis}
\label{sec:aba}
To further investigate agent behavior in UGC propagation, we analyze agent opinion scores and behavior types across posts of different popularity. As shown in Fig.~\ref{fig:beha} (a), when high-popularity posts are put into the social network simulation sandbox, an average of about 80\% of agents have an opinion score above 7.5, while only around 55\% of agents achieve the same score in low-popularity posts.
This indicates that high-popularity posts attract more agent attention, significantly raising their opinion scores. 
Additionally, agents exhibit more interactive behaviors (e.g., like, reply) with high-popularity posts.
This phenomenon aligns with real-world social network dynamics, where high-popularity content triggers more user engagement. 
This further validates the effectiveness of the PopSim, as it accurately models the propagation process in social networks, capturing the differentiated interest of agents toward posts of varying popularity, thereby improving SMPP performance.

\subsection{Impact of Agent Scale}
\label{sec:ias}
To examine the impact of agent scale on performance, we conduct experiments with agent scales of 10, 100, 200, 500, and 1000 on the SMPD dataset. As shown in Fig.~\ref{fig:impact} (a), the model's prediction performance improves progressively with increasing simulation scale. This suggests that larger simulation scales yield more realistic simulations, thereby enhancing SMPP performance. However, when the simulation scale reaches 200, the marginal benefits of further scale increases become less significant, while the simulation complexity increases substantially. Therefore, we consider a simulation scale of 100-200 as an optimal balance between efficiency and performance on this dataset. Nevertheless, more agents might be needed for a larger dataset. It can be discussed in future work.

\subsection{Impact of Interaction Rounds}
\label{sec:iir}
To explore the impact of the number of interaction simulation rounds on performance, we conduct experiments with interaction rounds ranging from 1 to 10. As shown in Fig.~\ref{fig:impact} (b), the model performs best when the number of interaction rounds is set to 6. This may be due to the fact that too few interaction rounds fail to adequately model UGC propagation and the deep interactions between users. On the other hand, too many interaction rounds tend to lead to opinion polarization in the final simulation results, reducing diversity and thus limiting the model's performance.

\begin{table}[!t]
\centering
\caption{Simulation efficiency. ``Time Cost / Step'' refers to the time required to complete a single simulation step, while ``avg. Total Time Cost'' denotes the average time to complete the entire UGC propagation simulation process. }
\label{tab:eff}
\resizebox{\linewidth}{!}{
\begin{tabular}{c|ccc}
\toprule
\textbf{Model} & \textbf{Time Cost / Step} & \textbf{avg. Total Time Cost} \\
\midrule
\textbf{Standard Simulation}    & 39.6s-89.2s           & 261.2s             \\
\textbf{Ours}    & 7.1s-18.6s          & 43.8s                             \\
\bottomrule
\end{tabular}
}
\end{table}

\subsection{Efficiency Analysis}
\label{sec:ea}
To validate the efficiency improvement of our simulation, we compared it with the standard simulation process using a system of 200 agents. In the standard simulation, agents interact with all their neighbors, whereas in the PopSim, we use the SMF-based agent interaction mechanism for simulation. The efficiency comparisons are presented in Table~\ref{tab:eff}. 
From the table, we can find that the SMF-based agent interaction reduces the average execution time by 83.2\% compared to the standard simulation. 
This result highlights the effectiveness of the proposed agent interaction mechanism in improving simulation efficiency.
\subsection{Case Study}
\label{sec:cs}
In this experiment, we simulate the propagation of a high-popularity post in the social network sandbox, as illustrated in Fig.~\ref{fig:case}. 
Initially, the agents in the network show little interest in the UGC. 
Direct popularity prediction may result in a low popularity score. 
After several rounds of interaction, agents gradually start discussing the UGC.
Eventually, the agents show high engagement in the propagation process, expressing interest through comments and posting related content. The final average opinion score of the agents is 8.1,
suggesting that the UGC may have high social media popularity in real-world scenarios.
This experiment demonstrates that PopSim can provide important clues for accurate popularity prediction through social network simulation.

\section{Conclusion}
In this work, we propose PopSim, a novel simulation-based SMPP paradigm.
Unlike the traditional inductive paradigm, PopSim leverages an LLM-based multi-agent social network sandbox to simulate UGC propagation dynamics for popularity prediction.
To accurately and efficiently simulate the UGC propagation in the network, we propose an SMF-based agent interaction mechanism, which models the dual-channel and bidirectional interactions between individuals and the population, enhancing agents' global perception and decision-making capabilities. 
Additionally, we design a multi-source information aggregation module to effectively integrate heterogeneous social metadata and propagation context, further enhancing SMPP performance.
Experiments on two real-world social media datasets demonstrate the effectiveness of PopSim, which outperforms existing SMPP models and achieves state-of-the-art performance. 
With the advancement of LLMs, PopSim can create more realistic simulations in the future, offering significant promise for research on social media analysis.

{
    \small
    \bibliographystyle{ieeenat_fullname}
    \bibliography{main}

@inproceedings{DBLP:conf/aaai/WuMCZ16,
  author       = {Bo Wu and
                  Tao Mei and
                  Wen{-}Huang Cheng and
                  Yongdong Zhang},
  title        = {Unfolding Temporal Dynamics: Predicting Social Media Popularity Using Multi-scale Temporal Decomposition},
  booktitle    = {{AAAI} Conference on Artificial Intelligence},
  pages        = {272--278},
  year         = {2016},
}

@inproceedings{DBLP:conf/mm/TanLLZZ22,
  author       = {Yunpeng Tan and
                  Fangyu Liu and
                  Bowei Li and
                  Zheng Zhang and
                  Bo Zhang},
  editor       = {Jo{\~{a}}o Magalh{\~{a}}es and
                  Alberto Del Bimbo and
                  Shin'ichi Satoh and
                  Nicu Sebe and
                  Xavier Alameda{-}Pineda and
                  Qin Jin and
                  Vincent Oria and
                  Laura Toni},
  title        = {An Efficient Multi-View Multimodal Data Processing Framework for Social
                  Media Popularity Prediction},
  booktitle    = {International Conference on Multimedia},
  pages        = {7200--7204},
  year         = {2022},
}

@inproceedings{DBLP:conf/icmcs/GuXTHHZ0G24,
  author       = {Jiayang Gu and
                  Xovee Xu and
                  Yulu Tian and
                  Yurun Hu and
                  Jiadong Huang and
                  Wenliang Zhong and
                  Fan Zhou and
                  Lianli Gao},
  title        = {{RRE:} {A} Relevance Relation Extraction Framework for Cross-domain
                  Recommender System at Alipay},
  booktitle    = {{IEEE} International Conference on Multimedia and Expo},
  pages        = {1--6},
  year         = {2024},
}

@inproceedings{DBLP:conf/nips/0001YL0020,
  author       = {Lei Bai and
                  Lina Yao and
                  Can Li and
                  Xianzhi Wang and
                  Can Wang},
  editor       = {Hugo Larochelle and
                  Marc'Aurelio Ranzato and
                  Raia Hadsell and
                  Maria{-}Florina Balcan and
                  Hsuan{-}Tien Lin},
  title        = {Adaptive Graph Convolutional Recurrent Network for Traffic Forecasting},
  booktitle    = {Advances in Neural Information Processing Systems},
  pages        = {17804--17815},
  year         = {2020},
}

@article{DBLP:journals/tomccap/XuWLWLZZL25,
  author       = {Ning Xu and
                  Xiaowen Wang and
                  Jing Liu and
                  Lanjun Wang and
                  Xuanya Li and
                  Mengxiao Zhu and
                  Yongdong Zhang and
                  An{-}An Liu},
  title        = {Model Can Be Subtle: Two Important Mechanisms for Social Media Popularity
                  Prediction},
  journal      = {{ACM} Trans. Multim. Comput. Commun. Appl.},
  volume       = {21},
  number       = {2},
  pages        = {71:1--71:20},
  year         = {2025},
}

@inproceedings{DBLP:conf/aaai/ZhangLCXLZZ24,
  author       = {Jienan Zhang and
                  Jie Liu and
                  Zhangtao Cheng and
                  Xovee Xu and
                  Fang Liu and
                  Ting Zhong and
                  Kunpeng Zhang},
  editor       = {Michael J. Wooldridge and
                  Jennifer G. Dy and
                  Sriraam Natarajan},
  title        = {THGFormer: Time-Aware Hypergraph Learning for Multimodal Social Media
                  Popularity Prediction (Student Abstract)},
  booktitle    = {{AAAI} Conference on Artificial Intelligence},
  pages        = {23705--23706},
  year         = {2024},
}

@inproceedings{DBLP:conf/mm/ChenHYCHZZ22,
  author       = {Weilong Chen and
                  Chenghao Huang and
                  Weimin Yuan and
                  Xiaolu Chen and
                  Wenhao Hu and
                  Xinran Zhang and
                  Yanru Zhang},
  title        = {Title-and-Tag Contrastive Vision-and-Language Transformer for Social
                  Media Popularity Prediction},
  booktitle    = {International Conference on Multimedia},
  pages        = {7008--7012},
  year         = {2022},
}

@inproceedings{DBLP:conf/mm/LaiZZ20,
  author       = {Xin Lai and
                  Yihong Zhang and
                  Wei Zhang},
  editor       = {Chang Wen Chen and
                  Rita Cucchiara and
                  Xian{-}Sheng Hua and
                  Guo{-}Jun Qi and
                  Elisa Ricci and
                  Zhengyou Zhang and
                  Roger Zimmermann},
  title        = {HyFea: Winning Solution to Social Media Popularity Prediction for
                  Multimedia Grand Challenge 2020},
  booktitle    = {International Conference on Multimedia},
  pages        = {4565--4569},
  year         = {2020},
}

@inproceedings{DBLP:conf/mm/LiHHL18,
  author       = {Liuwu Li and
                  Sihong Huang and
                  Ziliang He and
                  Wenyin Liu},
  editor       = {Susanne Boll and
                  Kyoung Mu Lee and
                  Jiebo Luo and
                  Wenwu Zhu and
                  Hyeran Byun and
                  Chang Wen Chen and
                  Rainer Lienhart and
                  Tao Mei},
  title        = {An Effective Text-based Characterization Combined with Numerical Features
                  for Social Media Headline Prediction},
  booktitle    = {International Conference on Multimedia},
  pages        = {2003--2007},
  year         = {2018},
}

@article{zhang2024contrastive,
  author       = {Zhizhen Zhang and
                  Ruihong Qiu and
                  Xiaohui Xie},
  title        = {Contrastive Learning for Implicit Social Factors in Social Media Popularity
                  Prediction},
  journal      = {CoRR},
  volume       = {abs/2410.09345},
  year         = {2024},
}

@inproceedings{DBLP:conf/aaai/XuZ0S25,
  author       = {Xovee Xu and
                  Yifan Zhang and
                  Fan Zhou and
                  Jingkuan Song},
  editor       = {Toby Walsh and
                  Julie Shah and
                  Zico Kolter},
  title        = {Improving Multimodal Social Media Popularity Prediction via Selective
                  Retrieval Knowledge Augmentation},
  booktitle    = {Association for the Advancement of Artificial
                  Intelligence},
  pages        = {932--940},

  year         = {2025},
}

@inproceedings{DBLP:conf/mm/JiPRC23,
  author       = {Liya Ji and
                  Chan Ho Park and
                  Zhefan Rao and
                  Qifeng Chen},
  title        = {Neural Image Popularity Assessment with Retrieval-augmented Transformer},
  booktitle    = {International Conference on Multimedia},
  pages        = {2427--2436},
  year         = {2023},
}

@inproceedings{kang2019catboost,
  title={Catboost-based framework with additional user information for social media popularity prediction},
  author={Kang, Peipei and Lin, Zehang and Teng, Shaohua and Zhang, Guipeng and Guo, Lingni and Zhang, Wei},
  booktitle={International Conference on Multimedia},
  pages={2677--2681},
  year={2019}
}

@inproceedings{wu2022deeply,
  title={Deeply exploit visual and language information for social media popularity prediction},
  author={Wu, Jianmin and Zhao, Liming and Li, Dangwei and Xie, Chen-Wei and Sun, Siyang and Zheng, Yun},
  booktitle={International Conference on Multimedia},
  pages={7045--7049},
  year={2022}
}

@inproceedings{xu2020multimodal,
  title={Multimodal deep learning for social media popularity prediction with attention mechanism},
  author={Xu, Kele and Lin, Zhimin and Zhao, Jianqiao and Shi, Peicang and Deng, Wei and Wang, Huaimin},
  booktitle={International Conference on Multimedia},
  pages={4580--4584},
  year={2020}
}

@inproceedings{wu2019smp,
  title={Smp challenge: An overview of social media prediction challenge 2019},
  author={Wu, Bo and Cheng, Wen-Huang and Liu, Peiye and Liu, Bei and Zeng, Zhaoyang and Luo, Jiebo},
  booktitle={International Conference on Multimedia},
  pages={2667--2671},
  year={2019}
}

@inproceedings{socialparti,
  author       = {Mattia Gasparini and
                  Robert Claris{\'{o}} and
                  Marco Brambilla and
                  Jordi Cabot},
  editor       = {Gregorio Robles and
                  Klaas{-}Jan Stol and
                  Xiaofeng Wang},
  title        = {Participation Inequality and the 90-9-1 Principle in Open Source},
  booktitle    = {International Symposium on Open Collaboration},
  pages        = {6:1--6:7},
  year         = {2020},
}

@article{lasry2007mean,
  title={Mean field games},
  author={Lasry, Jean-Michel and Lions, Pierre-Louis},
  journal={Japanese journal of mathematics},
  volume={2},
  number={1},
  pages={229--260},
  year={2007},
  publisher={Springer}
}

@article{DBLP:journals/advcs/DeffuantNAW00,
  author       = {Guillaume Deffuant and
                  David Neau and
                  Fr{\'{e}}d{\'{e}}ric Amblard and
                  G{\'{e}}rard Weisbuch},
  title        = {Mixing beliefs among interacting agents},
  journal      = {Adv. Complex Syst.},
  volume       = {3},
  number       = {1-4},
  pages        = {87--98},
  year         = {2000},
}

@inproceedings{DBLP:conf/iciap/OrtisFB19,
  author       = {Alessandro Ortis and
                  Giovanni Maria Farinella and
                  Sebastiano Battiato},
  editor       = {Elisa Ricci and
                  Samuel Rota Bul{\`{o}} and
                  Cees Snoek and
                  Oswald Lanz and
                  Stefano Messelodi and
                  Nicu Sebe},
  title        = {Prediction of Social Image Popularity Dynamics},
  booktitle    = {Image Analysis and Processing},
  series       = {Lecture Notes in Computer Science},
  volume       = {11752},
  pages        = {572--582},
  year         = {2019},
}

@inproceedings{DBLP:conf/www/KhoslaSH14,
  author       = {Aditya Khosla and
                  Atish Das Sarma and
                  Raffay Hamid},
  editor       = {Chin{-}Wan Chung and
                  Andrei Z. Broder and
                  Kyuseok Shim and
                  Torsten Suel},
  title        = {What makes an image popular?},
  booktitle    = {International World Wide Web Conference},
  pages        = {867--876},
  year         = {2014},
}

@inproceedings{DBLP:conf/mm/HsuLHT23,
  author       = {Chih{-}Chung Hsu and
                  Chia{-}Ming Lee and
                  Xiu{-}Yu Hou and
                  Chi{-}Han Tsai},
  editor       = {Abdulmotaleb El{-}Saddik and
                  Tao Mei and
                  Rita Cucchiara and
                  Marco Bertini and
                  Diana Patricia Tobon Vallejo and
                  Pradeep K. Atrey and
                  M. Shamim Hossain},
  title        = {Gradient Boost Tree Network based on Extensive Feature Analysis for
                  Popularity Prediction of Social Posts},
  booktitle    = {International Conference on Multimedia},
  pages        = {9451--9455},
  year         = {2023},
}

@article{DBLP:journals/tmm/XieZC23,
  author       = {Jiayi Xie and
                  Yaochen Zhu and
                  Zhenzhong Chen},
  title        = {Micro-Video Popularity Prediction Via Multimodal Variational Information
                  Bottleneck},
  journal      = {{IEEE} Trans. Multim.},
  volume       = {25},
  pages        = {24--37},
  year         = {2023},
  url          = {https://doi.org/10.1109/TMM.2021.3120537},
  doi          = {10.1109/TMM.2021.3120537},
  bibsource    = {dblp computer science bibliography, https://dblp.org}
}

@inproceedings{DBLP:conf/icml/0001LXH22,
  author       = {Junnan Li and
                  Dongxu Li and
                  Caiming Xiong and
                  Steven C. H. Hoi},
  editor       = {Kamalika Chaudhuri and
                  Stefanie Jegelka and
                  Le Song and
                  Csaba Szepesv{\'{a}}ri and
                  Gang Niu and
                  Sivan Sabato},
  title        = {{BLIP:} Bootstrapping Language-Image Pre-training for Unified Vision-Language
                  Understanding and Generation},
  booktitle    = {International Conference on Machine Learning},
  series       = {Proceedings of Machine Learning Research},
  volume       = {162},
  pages        = {12888--12900},
  year         = {2022},
}

@inproceedings{DBLP:conf/sigir/ZhongLZCZ024,
  author       = {Ting Zhong and
                  Jian Lang and
                  Yifan Zhang and
                  Zhangtao Cheng and
                  Kunpeng Zhang and
                  Fan Zhou},
  editor       = {Grace Hui Yang and
                  Hongning Wang and
                  Sam Han and
                  Claudia Hauff and
                  Guido Zuccon and
                  Yi Zhang},
  title        = {Predicting Micro-video Popularity via Multi-modal Retrieval Augmentation},
  booktitle    = {International Conference on
                  Research and Development in Information Retrieval},
  pages        = {2579--2583},
  year         = {2024},
}

@inproceedings{DBLP:conf/kdd/ChengZXTZ024,
  author       = {Zhangtao Cheng and
                  Jienan Zhang and
                  Xovee Xu and
                  Goce Trajcevski and
                  Ting Zhong and
                  Fan Zhou},
  title        = {Retrieval-Augmented Hypergraph for Multimodal Social Media Popularity
                  Prediction},
  booktitle    = {Conference on Knowledge Discovery
                  and Data Mining},
  pages        = {445--455},
  year         = {2024},
}

@article{touvron2023llama,
  author       = {Hugo Touvron and
                  Thibaut Lavril and
                  Gautier Izacard and
                  Xavier Martinet and
                  Marie{-}Anne Lachaux and
                  Timoth{\'{e}}e Lacroix and
                  Baptiste Rozi{\`{e}}re and
                  Naman Goyal and
                  Eric Hambro and
                  Faisal Azhar and
                  Aur{\'{e}}lien Rodriguez and
                  Armand Joulin and
                  Edouard Grave and
                  Guillaume Lample},
  title        = {LLaMA: Open and Efficient Foundation Language Models},
  journal      = {CoRR},
  volume       = {abs/2302.13971},
  year         = {2023},
}

@inproceedings{DBLP:conf/www/ZhangWWZ18,
  author       = {Wei Zhang and
                  Wen Wang and
                  Jun Wang and
                  Hongyuan Zha},
  editor       = {Pierre{-}Antoine Champin and
                  Fabien Gandon and
                  Mounia Lalmas and
                  Panagiotis G. Ipeirotis},
  title        = {User-guided Hierarchical Attention Network for Multi-modal Social
                  Image Popularity Prediction},
  booktitle    = {International Conference of World Wide Web},
  pages        = {1277--1286},
  year         = {2018},
}

@inproceedings{feng2019hypergraph,
  title={Hypergraph neural networks},
  author={Feng, Yifan and You, Haoxuan and Zhang, Zizhao and Ji, Rongrong and Gao, Yue},
  booktitle={Proceedings of the AAAI conference on artificial intelligence},
  volume={33},
  number={01},
  pages={3558--3565},
  year={2019}
}

@inproceedings{DBLP:conf/mm/WangWCHMZ20,
  author       = {Kai Wang and
                  Penghui Wang and
                  Xin Chen and
                  Qiushi Huang and
                  Zhendong Mao and
                  Yongdong Zhang},
  editor       = {Chang Wen Chen and
                  Rita Cucchiara and
                  Xian{-}Sheng Hua and
                  Guo{-}Jun Qi and
                  Elisa Ricci and
                  Zhengyou Zhang and
                  Roger Zimmermann},
  title        = {A Feature Generalization Framework for Social Media Popularity Prediction},
  booktitle    = {International Conference on Multimedia},
  pages        = {4570--4574},
  year         = {2020},
}

@inproceedings{weissburg2022judging,
  author       = {Evan Weissburg and
                  Arya Kumar and
                  Paramveer S. Dhillon},
  editor       = {Ceren Budak and
                  Meeyoung Cha and
                  Daniele Quercia},
  title        = {Judging a Book by Its Cover: Predicting the Marginal Impact of Title
                  on Reddit Post Popularity},
  booktitle    = {International Conference on Web
                  and Social Media},
  pages        = {1098--1108},
  year         = {2022},
}

@inproceedings{DBLP:conf/uist/ParkOCMLB23,
  author       = {Joon Sung Park and
                  Joseph C. O'Brien and
                  Carrie Jun Cai and
                  Meredith Ringel Morris and
                  Percy Liang and
                  Michael S. Bernstein},
  editor       = {Sean Follmer and
                  Jeff Han and
                  J{\"{u}}rgen Steimle and
                  Nathalie Henry Riche},
  title        = {Generative Agents: Interactive Simulacra of Human Behavior},
  booktitle    = {{ACM} Symposium on User Interface Software
                  and Technology},
  pages        = {2:1--2:22},

  year         = {2023},
}

@inproceedings{DBLP:conf/ijcai/LiuCZG0024,
  author       = {Yuhan Liu and
                  Xiuying Chen and
                  Xiaoqing Zhang and
                  Xing Gao and
                  Ji Zhang and
                  Rui Yan},
  title        = {From Skepticism to Acceptance: Simulating the Attitude Dynamics Toward
                  Fake News},
  booktitle    = {International Joint Conference on
                  Artificial Intelligence},
  pages        = {7886--7894},
  year         = {2024},
}

@article{DBLP:journals/corr/abs-2412-09237,
  author       = {Yijun Liu and
                  Wu Liu and
                  Xiaoyan Gu and
                  Yong Rui and
                  Xiaodong He and
                  Yongdong Zhang},
  title        = {LMAgent: {A} Large-scale Multimodal Agents Society for Multi-user
                  Simulation},
  journal      = {CoRR},
  volume       = {abs/2412.09237},
  year         = {2024},
}

@inproceedings{DBLP:conf/acl/MouWH24,
  author       = {Xinyi Mou and
                  Zhongyu Wei and
                  Xuanjing Huang},
  editor       = {Lun{-}Wei Ku and
                  Andre Martins and
                  Vivek Srikumar},
  title        = {Unveiling the Truth and Facilitating Change: Towards Agent-based Large-scale
                  Social Movement Simulation},
  booktitle    = {Association for Computational Linguistics},
  pages        = {4789--4809},
  year         = {2024},
}

@article{DBLP:journals/pami/DongSLG25,
  author       = {Xue Dong and
                  Xuemeng Song and
                  Tongliang Liu and
                  Weili Guan},
  title        = {Prompt-Based Multi-Interest Learning Method for Sequential Recommendation},
  journal      = {{IEEE} Trans. Pattern Anal. Mach. Intell.},
  volume       = {47},
  number       = {8},
  pages        = {6876--6887},
  year         = {2025},
}

@inproceedings{DBLP:conf/kdd/LiLMWP15,
  author       = {Cheng Li and
                  Yue Lu and
                  Qiaozhu Mei and
                  Dong Wang and
                  Sandeep Pandey},
  editor       = {Longbing Cao and
                  Chengqi Zhang and
                  Thorsten Joachims and
                  Geoffrey I. Webb and
                  Dragos D. Margineantu and
                  Graham Williams},
  title        = {Click-through Prediction for Advertising in Twitter Timeline},
  booktitle    = {International Conference on Knowledge Discovery and Data Mining},
  pages        = {1959--1968},
  year         = {2015},
}

@article{DBLP:journals/tnsm/DAlconzoDMMC19,
  author       = {Alessandro D'Alconzo and
                  Idilio Drago and
                  Andrea Morichetta and
                  Marco Mellia and
                  Pedro Casas},
  title        = {A Survey on Big Data for Network Traffic Monitoring and Analysis},
  journal      = {{IEEE} Trans. Netw. Serv. Manag.},
  volume       = {16},
  number       = {3},
  pages        = {800--813},
  year         = {2019},
}

@article{liu2025rumorsphere,
  author       = {Yijun Liu and
                  Wu Liu and
                  Xiaoyan Gu and
                  Weiping Wang and
                  Jiebo Luo and
                  Yong{-}Dong Zhang},
  title        = {RumorSphere: {A} Framework for Million-scale Agent-based Dynamic Simulation
                  of Rumor Propagation},
  journal      = {CoRR},
  volume       = {abs/2509.02172},
  year         = {2025},
}

@inproceedings{DBLP:conf/hicss/BrunkerWMM20,
  author       = {Felix Br{\"{u}}nker and
                  Magdalena Wischnewski and
                  Milad Mirbabaie and
                  Judith Meinert},
  title        = {The Role of Social Media during Social Movements - Observations from
                  the {\#}metoo Debate on Twitter},
  booktitle    = {Hawaii International Conference on System Sciences},
  pages        = {1--10},
  year         = {2020},
}

@inproceedings{DBLP:conf/mir/CappalloMS15,
  author       = {Spencer Cappallo and
                  Thomas Mensink and
                  Cees G. M. Snoek},
  title        = {Latent Factors of Visual Popularity Prediction},
  booktitle    = {International Conference on Multimedia Retrieval},
  pages        = {195--202},
  year         = {2015},
}

@inproceedings{DBLP:conf/ijcai/WuCZHLM17,
  author       = {Bo Wu and
                  Wen{-}Huang Cheng and
                  Yongdong Zhang and
                  Qiushi Huang and
                  Jintao Li and
                  Tao Mei},
  editor       = {Carles Sierra},
  title        = {Sequential Prediction of Social Media Popularity with Deep Temporal
                  Context Networks},
  booktitle    = {International Joint Conference on
                  Artificial Intelligence},
  pages        = {3062--3068},
  year         = {2017},
}

@article{DBLP:journals/ipm/QianXLLJCL22,
  author       = {Yang Qian and
                  Wang Xu and
                  Xiao Liu and
                  Haifeng Ling and
                  Yuanchun Jiang and
                  Yidong Chai and
                  Ye{-}Zheng Liu},
  title        = {Popularity prediction for marketer-generated content: {A} text-guided
                  attention neural network for multi-modal feature fusion},
  journal      = {Inf. Process. Manag.},
  volume       = {59},
  number       = {4},
  pages        = {102984},
  year         = {2022},
}

@inproceedings{DBLP:conf/kdd/JiLLSL0023,
  author       = {Shuo Ji and
                  Xiaodong Lu and
                  Mingzhe Liu and
                  Leilei Sun and
                  Chuanren Liu and
                  Bowen Du and
                  Hui Xiong},
  editor       = {Ambuj K. Singh and
                  Yizhou Sun and
                  Leman Akoglu and
                  Dimitrios Gunopulos and
                  Xifeng Yan and
                  Ravi Kumar and
                  Fatma Ozcan and
                  Jieping Ye},
  title        = {Community-based Dynamic Graph Learning for Popularity Prediction},
  booktitle    = {Conference on Knowledge Discovery and Data Mining},
  pages        = {930--940},
  year         = {2023},
}

@inproceedings{DBLP:conf/cvpr/DongLCPZ24,
  author       = {Zhikang Dong and
                  Xiulong Liu and
                  Bin Chen and
                  Pawel Polak and
                  Peng Zhang},
  title        = {MuseChat: {A} Conversational Music Recommendation System for Videos},
  booktitle    = {{IEEE/CVF} Conference on Computer Vision and Pattern Recognition},
  pages        = {12775--12785},
  year         = {2024},
}

@inproceedings{DBLP:conf/cvpr/LiFYDTS23,
  author       = {Hao Li and
                  Charless C. Fowlkes and
                  Hao Yang and
                  Onkar Dabeer and
                  Zhuowen Tu and
                  Stefano Soatto},
  title        = {Guided Recommendation for Model Fine-Tuning},
  booktitle    = {{IEEE/CVF} Conference on Computer Vision and Pattern Recognition},
  pages        = {3633--3642},
  year         = {2023},
}

@inproceedings{DBLP:conf/cvpr/LiX0CJN25,
  author       = {Zaijing Li and
                  Yuquan Xie and
                  Rui Shao and
                  Gongwei Chen and
                  Dongmei Jiang and
                  Liqiang Nie},
  title        = {Optimus-2: Multimodal Minecraft Agent with Goal-Observation-Action
                  Conditioned Policy},
  booktitle    = {{IEEE/CVF} Conference on Computer Vision and Pattern Recognition},
  pages        = {9039--9049},
  year         = {2025},
}

@article{DBLP:journals/corr/abs-2407-21783,
  author       = {Llama Team},
  title        = {The Llama 3 Herd of Models},
  journal      = {CoRR},
  volume       = {abs/2407.21783},
  year         = {2024},
}

@article{DBLP:journals/corr/abs-2504-10157,
  author       = {Xinnong Zhang and
                  Jiayu Lin and
                  Xinyi Mou and
                  Shiyue Yang and
                  Xiawei Liu and
                  Libo Sun and
                  Hanjia Lyu and
                  Yihang Yang and
                  Weihong Qi and
                  Yue Chen and
                  Guanying Li and
                  Ling Yan and
                  Yao Hu and
                  Siming Chen and
                  Yu Wang and
                  Xuanjing Huang and
                  Jiebo Luo and
                  Shiping Tang and
                  Libo Wu and
                  Baohua Zhou and
                  Zhongyu Wei},
  title        = {SocioVerse: {A} World Model for Social Simulation Powered by {LLM}
                  Agents and {A} Pool of 10 Million Real-World Users},
  journal      = {CoRR},
  volume       = {abs/2504.10157},
  year         = {2025},
}

@article{DBLP:journals/corr/abs-2410-04360,
  author       = {Jiakai Tang and
                  Heyang Gao and
                  Xuchen Pan and
                  Lei Wang and
                  Haoran Tan and
                  Dawei Gao and
                  Yushuo Chen and
                  Xu Chen and
                  Yankai Lin and
                  Yaliang Li and
                  Bolin Ding and
                  Jingren Zhou and
                  Jun Wang and
                  Ji{-}Rong Wen},
  title        = {GenSim: {A} General Social Simulation Platform with Large Language
                  Model based Agents},
  journal      = {CoRR},
  volume       = {abs/2410.04360},
  year         = {2024},
}

@inproceedings{DBLP:conf/cvpr/McKeeSSR23,
  author       = {Daniel McKee and
                  Justin Salamon and
                  Josef Sivic and
                  Bryan C. Russell},
  title        = {Language-Guided Music Recommendation for Video via Prompt Analogies},
  booktitle    = {{IEEE/CVF} Conference on Computer Vision and Pattern Recognition},
  pages        = {14784--14793},
  year         = {2023},
}
}

\end{document}